# Energy-dependent broadening of Rashba-type spin splitting in a $Au_2Sb$ surface alloy with periodic structural defects

Jinbang Hu[1,2 +]; Xiansi Wang[2]; Anna Cecilie Åsland[1]; Justin W. Wells[1,3 +]

[1]*Department of Physics, Norwegian University of Science and Technology (NTNU), NO-7491, Trondheim, Norway.*

[2]*Hunan University, Changsha, 410082, China.*

[3]*Semiconductor Physics, Department of Physics, University of Oslo (UiO), NO-0371 Oslo, Norway.*

[+]Corresponding author: hujinbang1@gmail.com; j.w.wells@fys.uio.no.

## Abstract

Here, we report a novel two-dimensional (2D) $Au_2Sb$ superstructure on Au(111) that exhibits both consistencies and discrepancies with the expected electronic features of an ideal 2D surface alloy with $\sqrt{3} \times \sqrt{3}$ periodicity. Using spin- and angle-resolved photoemission spectroscopy, we observe a spin splitting of the alloy bands with antiparallel spin polarization, stemming from Rashba spin-orbit coupling. Notably, the Rashba bands are significantly broadened in comparison to idealized expectations. Taking advantage of the good agreement between the experimental results and DFT calculations, we identify the broadening of the Rashba bands as a consequence of perturbations introduced by three-pointed-star-shaped defects, which act as nonresonant impurities within the $Au_2Sb$ superstructure. These periodic defects are capable of shifting the energy position of the Rashba bands without breaking the in-plane rotational and mirror symmetries, which suggests that the deliberate introduction of periodic defects into a Rashba SOC system possesses great potential in engineering the spin-dependent properties of spintronic devices.

## Introduction

Since the experimental discovery of 2D graphene, the investigation of 2D materials has attracted significant attention[1-3]. Plenty of efforts have been made to improve the tunability and performance of various 2D materials[4-6]. The exfoliation of single-layer black phosphorous opened a door to the investigation of VA-element 2D materials with excellent properties[7]. Compared to mechanical exfoliation, molecular beam epitaxial (MBE) growth of 2D materials is more controllable and flexible for different systems. Among VA elements, bismuthene has been successfully manufactured by MBE on SiC[8] and Ag[9], exhibiting exotic topological features. During the growth of mono-elemental 2D materials on metals (such as Cu, Ag, and Au), the atoms of the sample material are adsorbed onto the substrate and form a surface alloy at low coverage due to the strong interfacial interaction. The surface alloy may also possess interesting properties[10].

One reason why 2D materials are interesting is the possibility of exhibiting Rashba spin-orbit coupling (SOC) effect[11], which exists in inversion-symmetry broken systems. Rashba SOC is related to rich physical phenomena and can break the degeneracy of the spin degree of freedom without any external field or simultaneous magnetization, making it useful for spintronics applications. Materials of VA elements, such as Bi[12-14] and Sb[15,16] are also a versatile playground for the research on Rashba physics. In this paper, we introduce a $Au_2Sb$ surface alloy decorated with periodic defects. The 3-pointed-star-shaped defects preserve the in-plane rotational symmetry and the mirror symmetry. The presence of the spin splitting of the alloy bands with antiparallel spin polarization is demonstrated by spin- and angle-resolved photoemission spectroscopy (ARPES). The good agreement between density functional theory (DFT) calculations and the experimental results confirms that the periodically distributed structural defects in the $Au_2Sb$ surface alloy act as nonresonant impurities, and that the Rashba bands can be shifted by properly introducing these defects. Our results indicate that the periodic nonresonant impurities serve as an additional way to tune the Rashba bands.

## Results

### Structure characterization

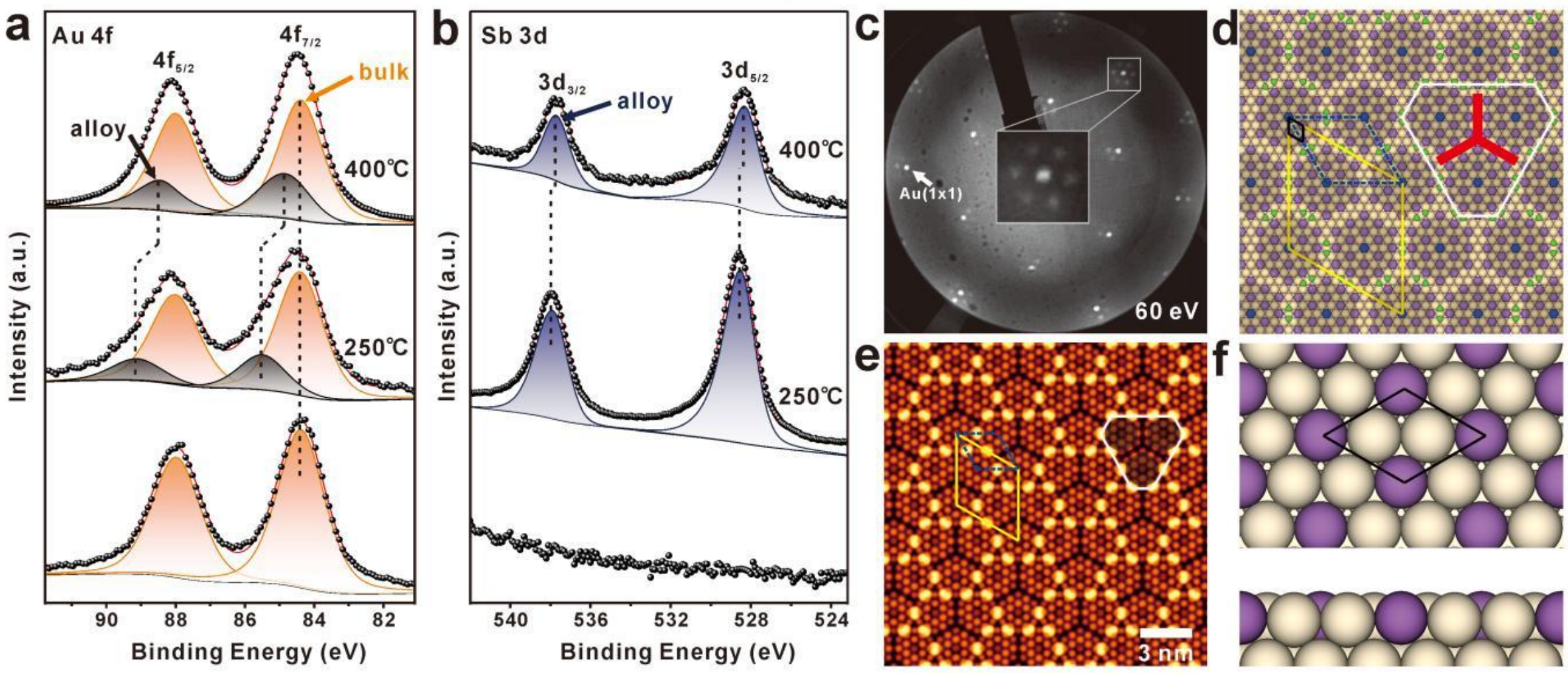

**Figure 1. Characterization of the moiré superstructure formed by Sb atoms alloying on the Au surface.** (a-b) XPS spectra of Au 4*f* (a) and Sb 3*d* (b) core levels taken before growth, for the Sb film after growth and annealing at 250 °C, and after further annealing at 400 °C. (c) LEED pattern (60 eV) from the $Au_2Sb$ surface alloy. The inset zooms in the diffraction around the Au(1x1) diffraction spot, presenting a $8 \times 8$ moiré superstructure. (d) Energetically favourable atomic model of the surface alloy with the moiré superstructure marked by the gray hexagon-shaped shadow. (e) STM simulation of the moiré superstructure with Sb dimers adsorbed above the $Au_2Sb$ surface alloy. The positions of adsorbed Sb dimers around the sides of the quasi-hexagon lattice are marked by green balls in (d). The yellow rhombus and blue dashed rhombus overlaid in (d-e) demonstrate the orientation and unit cell of the quasi-hexagon phase [marked by white lines in (d-e)] assembled by Sb dimers and the $8 \times 8$ moiré superstructure, respectively. The light yellow balls are Au atoms, and balls of other colors are Sb atoms. A 3-pointed-star-shaped defect is marked by the thick red lines. (f) Top and side view of the fully optimized slab lattice of the $Au_2Sb$ surface alloy considered for band structure discussion.

Figure 1 presents the $Au_2Sb$ surface alloy fabricated on Au(111) at approximately 400 °C. XPS measurements were conducted to monitor the chemical shift of the Au 4*f* and Sb 3*d* core levels during the phase transition process, commencing from a well-prepared 2D Sb(110) rhombohedral structure followed by annealing up to 400 °C. The initial 2D Sb(110) rhombohedral phase on the Au(111) surface was prepared by annealing a thicker Sb film sample, as reported in Ref.[17], and confirmed by the LEED pattern (60 eV) in Supplementary Figure 2a. It is obvious in Figure 1b that the Sb 3*d* spectrum of the as-fabricated 2D Sb(110) rhombohedral phase at 250 °C exhibits two characteristic components, denoted as Sb $3d_{3/2}$ and $3d_{5/2}$, with energies of 537.95 eV and 528.60 eV, respectively. Compared to the Au 4*f* signal of clean Au(111), these Au 4*f* peaks in Figure 1a show a small systematic shift towards the higher binding energy side by approximately 0.15 eV, and also deform in shape after the growth of the Sb monolayer. The fit of the Au 4*f* signal reveals a remaining bulk component (orange) and a surface component (black) at higher binding energy shifting towards the bulk component by 1.09 eV relative to clean Au(111), primarily due to the interaction with the Sb adlayer. Subsequently, annealing at 400 °C induced a change in the shape of the Au 4*f* peak. The peak fitting indicated that the surface component shifted by 0.50 eV towards the bulk component when annealing from 250 °C to 400 °C. Additionally, the Sb 3d peak shifted towards the lower binding energy side by 0.20 eV. This observation suggests that an Au-Sb surface alloy might be fabricated due to the altered interaction between the Au surface and the Sb adlayer. To further investigate the surface structure of Sb on Au(111) after annealing at 400 °C, we obtained a LEED pattern (beam energy 60 eV) of a $\sqrt{3} \times \sqrt{3}R30^{\circ}$ structure , as shown in Figure 1c. A clear moiré pattern surrounding the Au(1x1) diffraction spot is observed in the zoomed-in LEED image, compared to that of the clean Au(111) surface. This observation signifies a phase transition from the monolayer Sb(110) rhombohedral phase to an $8 \times 8$ superstructure. The formation of the $8 \times 8$ superstructure aligns well with the analysis based on atomically resolved STM images presented in Refs.[17,18] (see Supplementary Figures 4c and 5c in Ref.[17], and Figure 3b in Ref.[18] for details).

To gain further insights into the experimental findings, we employed DFT calculations to construct an optimized $8 \times 8$ structural model of the $Au_2Sb$ surface alloy, as depicted in Figure 1d. The 3-pointed-star-shaped defects (indicated by thick red lines) separate the moiré superstructure (highlighted by the gray hexagon-shaped shadow) formed by alloying of Au and Sb atoms. We then performed STM simulations based on this model, incorporating Sb dimers, as shown in Figure 1e. Remarkably, the simulated STM image accurately reproduces both the moiré features and the quasi-hexagonal lattice observed in previous experiments[18]. Notably, 12 equidistant dimers form a quasi-hexagon with three short sides and three long sides, originating from the adsorbed Sb dimers above the $Au_2Sb$ surface alloy. The Sb dimers act as an intermediate state in the alloying-to-dealloying transition as the Sb coverage increases to form a rhombohedral phase at approximately one monolayer.

Additionally, we found that the same LEED pattern with a moiré structure can be produced by directly depositing a submonolayer of Sb on the Au(111) surface at room temperature (see Supplementary Figure 2 for details). In this case, the XPS of the Sb 3d core level has two components, separated by 0.76 eV. The component at the higher binding energy side corresponds to the contribution from Sb dimers assembled in a quasi-hexagon phase, as reported in Ref.[18] Compared to the $Au_2Sb$ surface alloy fabricated by depositing a submonolayer of Sb at room temperature, annealing the 2D Sb(110) rhombohedral phase demonstrates superior performance in achieving full alloying between Au and Sb, effectively avoiding the formation of various assembled phases by Sb dimers above the $Au_2Sb$ surface alloy[18].

## Confirmation of band splitting

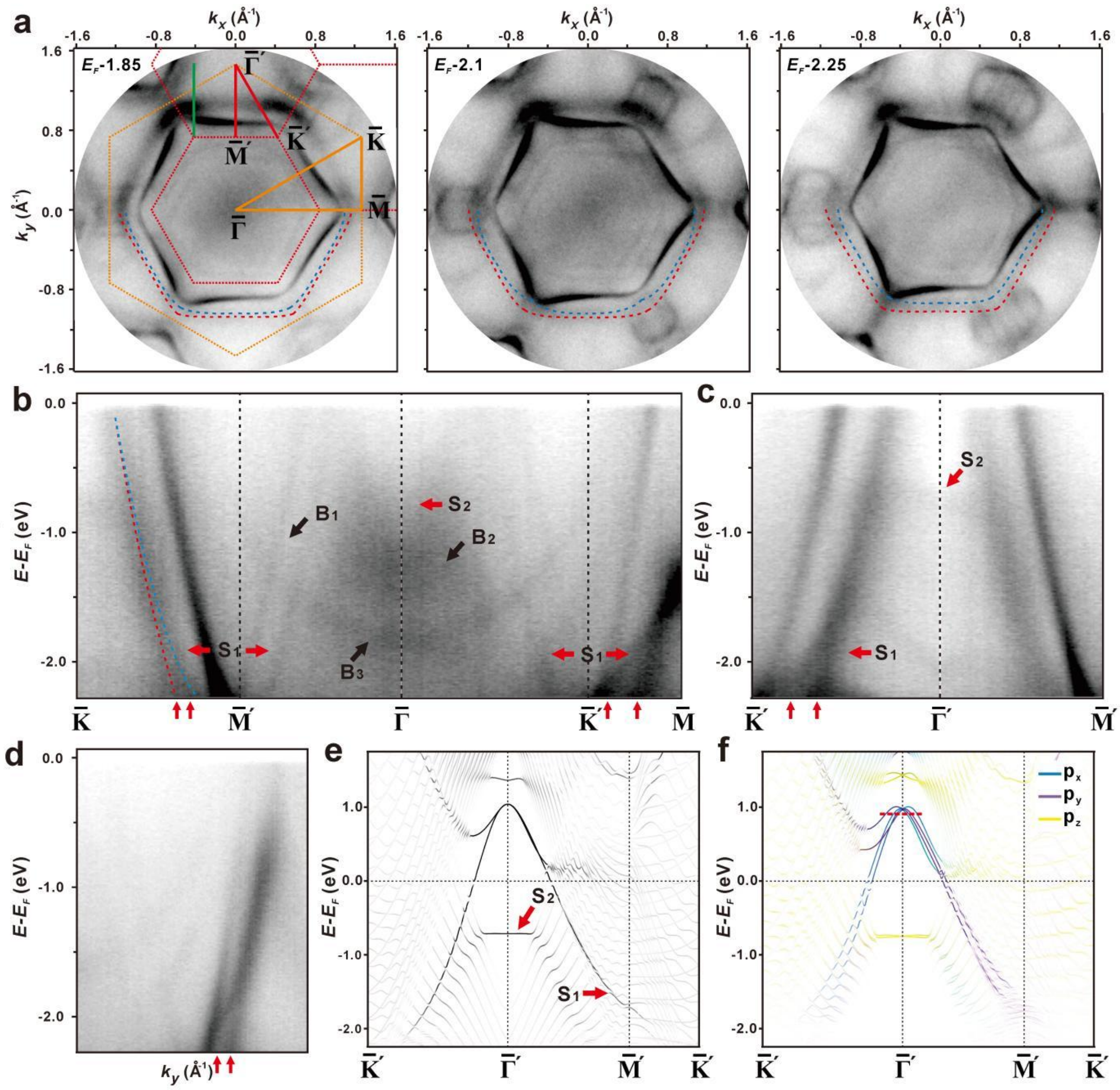

**Figure 2. Electronic structure of the $Au_2Sb$ surface alloy.** (a) ARPES data of constant energy contours throughout the SBZ at different binding energies. $1 \times 1$ (yellow) and $\sqrt{3} \times \sqrt{3}$ (red) SBZ with high symmetry point labels overlaid on the constant energy contours at 1.85 eV below the Fermi level. The red and blue dashed lines in (a) guide the tendency of the splitting of the band. (b) and (c) ARPES results along the $\overline{\mathrm{K}} - \overline{\Gamma} - \overline{\mathrm{M}}$ line of the Au(111) SBZ and $\overline{\mathrm{K}}' - \overline{\Gamma}' - \overline{\mathrm{M}}'$ line of the 2nd $\sqrt{3} \times \sqrt{3}$ SBZ of the $Au_2Sb$ surface alloy, respectively. A few observations of particular interest are indicated by arrows. (d) Energy–momentum slice along the green line marked in (a). (e-f) Calculated band structure without (e) and with (f) SOC for the model of the $\sqrt{3} \times \sqrt{3}$ $Au_2Sb$ surface alloy. Black symbols in (e) indicate contributions from the topmost $Au_2Sb$ layer. Blue, purple, and yellow symbols indicate contributions from $p_x$, $p_y$, and $p_z$ states, respectively, in (f). The red arrows at the bottom of (b-d) indicate the splitting character of $S_1$.

In the moiré superstructure formed by Sb atoms alloying on the Au surface, the incorporated Sb atoms are surrounded by a ligand environment similar to that of other $\sqrt{3} \times \sqrt{3}R30°$surface alloy structures[19-21]. To investigate the potential occurrence of Rashba splitting, we turn to the electronic band structure investigated by ARPES. Figures 2a-d present the band structures of the $Au_2Sb$ surface alloy. The constant energy contours in Figure 2a exhibit band features from both Au substrate and the $Au_2Sb$ surface alloy. Notably, a new captivating feature appears as a distorted hexagon that connects the $\overline{\mathrm{M}}$ points of the Au(111) SBZ. Upon closer inspection, the distorted hexagon band splits into two sub-contours, indicated by the red and blue dashed lines in Figure 2a. The splitting becomes more pronounced with increasing binding energy. Interestingly, the shapes of the distorted hexagon band resemble those reported for $Au_2Sn$[22,23], as can be compared with Supplementary Figures 2c-e in the Supplemental Information[24].

In Figure 2a, the $1 \times 1$ (yellow) and $\sqrt{3} \times \sqrt{3}$ (red) SBZs and their respective high symmetry points have been labelled and overlaid on the constant energy contour at 1.85 eV below the Fermi level ($E_b = 0$ eV). The band structure between the high symmetry points in the Au(111) SBZ and the second $\sqrt{3} \times \sqrt{3}$ SBZ are shown in Figures 2b and 2c, respectively. We observe the emergence of a new set of band features around the $\overline{\Gamma}$ point after Sb alloying on the Au surface. Specifically, the fuzzy $S_2$ band and the hole-like parabolically dispersing bands $S_1$ and $B_1$ appear to be periodic with respect to the $\sqrt{3} \times \sqrt{3}$ unit cell. The dispersing band $B_1$ arises from Umklapp scattering of the Au $sp$ band emission due to the presence of the $\sqrt{3} \times \sqrt{3}$ structure[23,25]. Additionally, the weak bands labelled $B_2$ and $B_3$ are observed in the first $\sqrt{3} \times \sqrt{3}$ SBZ but are absent in the second $\sqrt{3} \times \sqrt{3}$ SBZ. These bands represent replicas of the Au bulk band structure, resulting from photoexcitation by the He $I_\beta$ emission line of the nonmonochromatized He $I$ light source. Furthermore, all the bulk-related band features $B_1$-$B_3$ observed in Figure 2b have also been observed on Sn/Au(111)–$\sqrt{3} \times \sqrt{3}$ [22]. The band features we are mainly concerned about, $S_1$ and $S_2$, are mixed with the Au bulk bands, and look weak and fuzzy. However, they exhibit higher signals in the second $\sqrt{3} \times \sqrt{3}$ SBZ (Figure 2c). The splitting of the $S_1$ band into two sub-bands, observed along the $\overline{\Gamma}' - \overline{K}'$ direction (Figure 2c) and also in the slice (Figure 2d) along the green line marked in Figure 2a, becomes clear as the binding energy increases. The spin degeneracy of $S_1$ is lifted even further along the $\overline{K}' - \overline{M}$ direction below the binding energy $E_b = 1.5$ eV, see the right side in Figure 2b. The evolution of the splitting of the band feature $S_1$ along a direction normal to the $\overline{\Gamma} - \overline{\Gamma'}$ line is shown in Supplementary Figures 3a-b[24]. Additionally, the emission intensity of $S_2$ is too weak to reveal any clear dispersive features.

To get a better understanding of the band structure, DFT calculations with a well-ordered $\sqrt{3} \times \sqrt{3}R30°$ lattice (Figure 1f) were performed. At this stage, compared to the $8 \times 8$ superstructure concluded from LEED and STM data[18], the 3-pointed-star-shaped defects were ignored. This can be done since each Sb atom in both the ideal $\sqrt{3} \times \sqrt{3}R30°$ lattice and the superstructure is surrounded by 6 Au atoms as nearest ligands, and the in-plane symmetry (e.g., the rotational symmetry and in-plane inversion symmetry) of the $8 \times 8$ superstructure is the same as that of the $\sqrt{3} \times \sqrt{3}R30°$ lattice. Also, the $8 \times 8$ superstructure is too large for the numerical calculation. We will discuss the influence from the periodic defects at a later stage. The surface band structure without SOC and with SOC projected onto the topmost $Au_2Sb$ surface alloy layer are shown in Figures 2e and 2f, respectively. We find that the experimentally observed bands $S_1$ and $S_2$ can be clearly identified, and the inclusion of SOC induces an obvious spin splitting of the band $S_1$, displaying a clear separation of two spin sub-bands and a steep energy dispersion. Conversely, the calculated $S_2$ band, located at around $E_b = 0.75$ eV, exhibits a flat region near the $\overline{\Gamma}'$ point with quite weak splitting. The band structure calculated with SOC (Figure 2f) shows qualitative agreement with the experimentally observed bands $S_1$ and $S_2$. However, the calculated band structure from a model of Sb adsorbed above Au(111) with a $\sqrt{3} \times \sqrt{3}$ surface reconstruction (see Supplementary Figure 4a for details) exhibits distinct differences compared to the experimental ARPES data, indicating that Sb alloying on the Au surface forms a $8 \times 8$ superstructure. Further analysis of the orbital-projected band structures of the $s$, $p$, and $d$ orbitals (Figure 2f and Supplementary Figure 4b) indicates that band $S_1$ mainly consists of Sb $p_{\parallel}$ ( $p_{\parallel} = p_x, p_y$ ) and Au $d_{\parallel}$ ( $d_{\parallel} = d_{xy}, d_{x^2-y^2}$ ) orbitals, suggesting a direct hybridization between the Sb and Au bands. On the other hand, the $S_2$ hybrid band mainly contains Sb $p_z$ and Au $s$ orbital components (see Supplementary Figure 4b for details).

### Spin contrast in Rashba bands

One typical feature of Rashba SOC is the presence of a helical spin texture. With Rashba SOC, the spin degeneracy is removed, resulting in two sub-bands with opposite spin directions. To further unravel the spin texture of the hole-like parabolic surface band $S_1$, we performed spin-resolved ARPES measurements, filtering the in-plane spin component along the $\pm k_y$ direction. Spin filtering is realized by introducing a spin polarizing electron mirror into the electron optical path, where electrons with opposite spins experience different scattering potentials during the reflecting process[26,27]. It is found that the seldomly spin polarized Au $sp$ band shows no significant difference in the individual spin-filtered images (Figures 3a and 3b), indicating that there is an almost equal intensity of electrons with spin along $+ k_y$ and $- k_y$ in the Au $sp$ band. This means that the spin-degenerate band is effectively removed in the spin-resolved image (Figure 3c), which is obtained by taking the difference between the two spin-filtered images. Details on the calculation of the spin resolved data are provided in the method section. In contrast, the two branches of the Rashba-type $S_1$ band, highlighted by the red and blue dashed curves, exhibit subtle signs of spin contrast.

Figures 3d-f and 3g-i depict the band dispersion of the Rashba-type $S_1$ state along the green lines 1 and 2 marked in Figure 3c, respectively. These two lines are aligned similarly to the green line in the Au(111) SBZ shown in Figure 2a, where the two sub-bands of the $S_1$ band in the slice (Figure 2d) exhibit a clear and strong splitting. Meanwhile, the Au band appears with relatively weak intensity, ensuring that the $S_1$ band is not obscured by signals from the Au band. In the partial photoemission intensity maps in Figures 3d-e, it is apparent that the distance between the Au $sp$ band and the $S_1$ band is smaller in Figure 3d (filtering spin along $-k_y$ ) than in Figure 3e (filtering spin along $+k_y$ ). These differences in the measured projection of the spin along $\pm k_y$ are more clearly demonstrated in the spin-resolved image (Figure 3f), and also in the momentum distribution curves (MDCs) of the spin polarization at $E_b = 1.9$ eV (top panel of Figure 3j). Notably, the intensity profiles of the two sub-bands, marked by the green circle and black circle respectively, appear to be opposite. Moreover, Figures 3g-i present the analysis of the spin-resolved band dispersion along the green line 2 marked in Figure 3c, also revealing that the two branches of the $S_1$ band are spin-polarized.

A representative DFT calculated spin texture of the Rashba-type $S_1$ band in the $k_x k_y$ plane at 0.9 eV above the Fermi level is shown in Figure 3k, where the arrows indicate the in-plane orientation of the spin. A clear Rashba-type helical spin texture can be seen in the two branches of the $S_1$ band with opposite spin polarization, which aligns well with our observation from spin-resolved ARPES measurements.

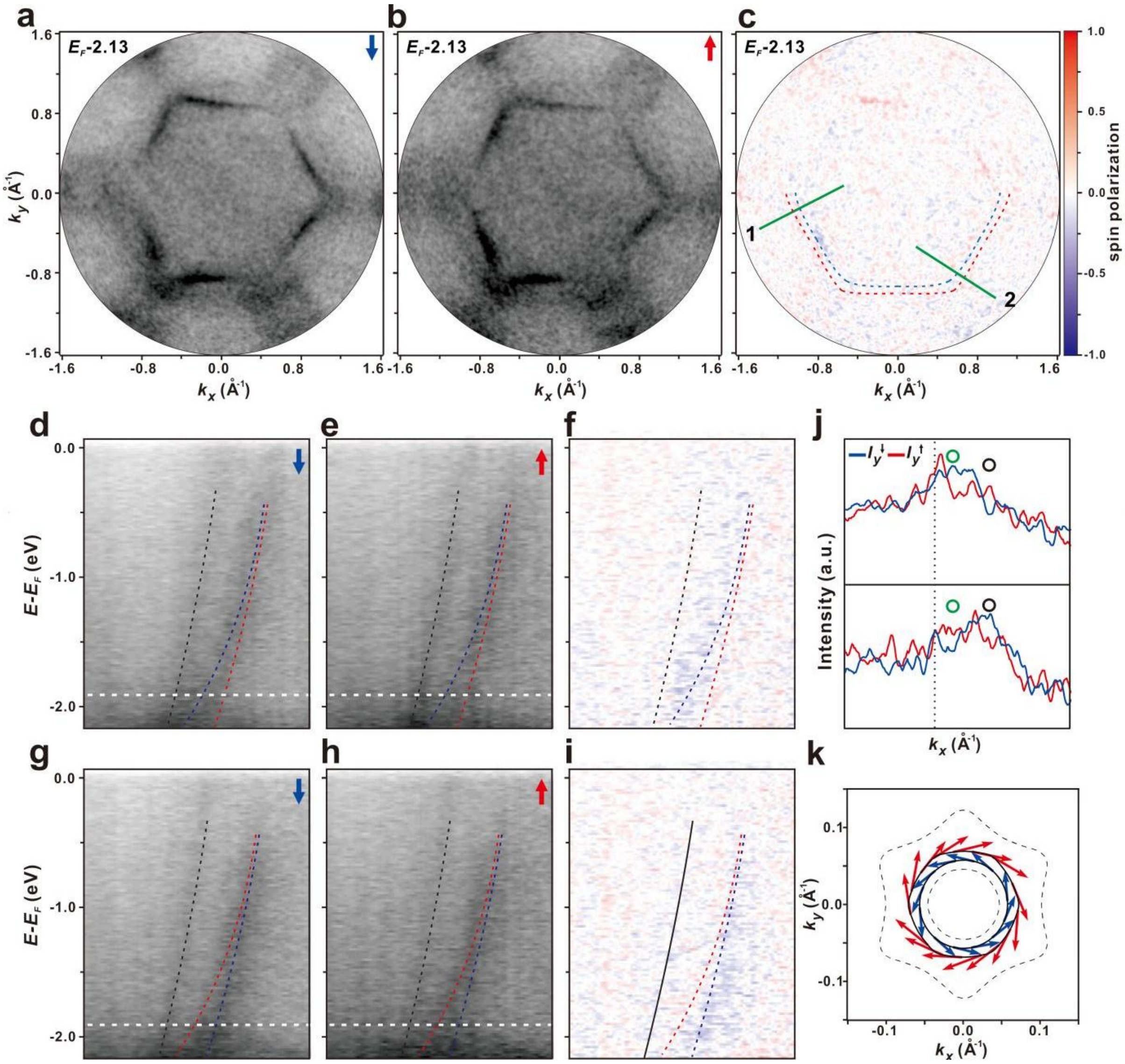

**Figure 3. Spin-resolved ARPES measurement of the Rashba-type band $S_1$ when filtering spin along $\pm k_y$.** (a-c) The constant energy surfaces were measured at $E_b = 2.13$ eV: (a) partial photoemission intensity map for the spin-polarized component along the $-k_y$ direction, (b) spin-polarized component along the $+k_y$ direction, (c) Spin resolved image. The color scale indicates the spin polarization $P_y$, from $P_y = -1$ (blue) to $P_y = 1$ (red). (d-f) and (g-i) are the same as (a-c) but for band dispersion of the Rashba-type $S_1$ band along the green lines 1 and 2 marked in (c), respectively. The dark line, blue line and red line used in (d-i) are used as references of the position of the Au $sp$ band and two sub-bands of $S_1$. (j) In-plane spin polarized MDCs along $k_x$ at $E_b = 1.9$ eV marked as white dashed line in (d-e) and (g-i), respectively. Red (blue) curve shows the spin-polarized component of the intensity along $+k_y(-k_y)$ direction. The black dashed line and circle markers denote the peak positions of the Au $sp$ band and two sub-bands of $S_1$ in the MDCs. (k) Calculated spin texture of the Rashba-type $S_1$ band in the $k_x k_y$ plane at $E = E_F + 0.9$ eV marked by the red dashed line in Figure 2(f). The length and the direction of the arrows indicate the spin magnitude and direction.

## Origin of the broadening character in Rashba bands

The Rashba-type SOC has been widely reported in similar systems, such as the $Au_2Sn$ surface alloy[22,23]. The pronounced Rashba-type band feature predicted by theoretical calculations in the $Au_2Sb$ surface alloy appears clearly in our experimental results, but it exhibits a broadening character in the two sub-bands. This discrepancy may be attributed to the periodic defects that were ignored in the current theoretical consideration. To gain further insight into the origin of the broadening character observed in the band feature $S_1$ from the $Au_2Sb$ surface alloy, we introduced perturbations to the $Au_2Sb$ surface alloy by subsequently substituting the Sb atoms with Au atoms as point defects in an $2 \times 2$ reconstruction model for theoretical study (Figure 4).

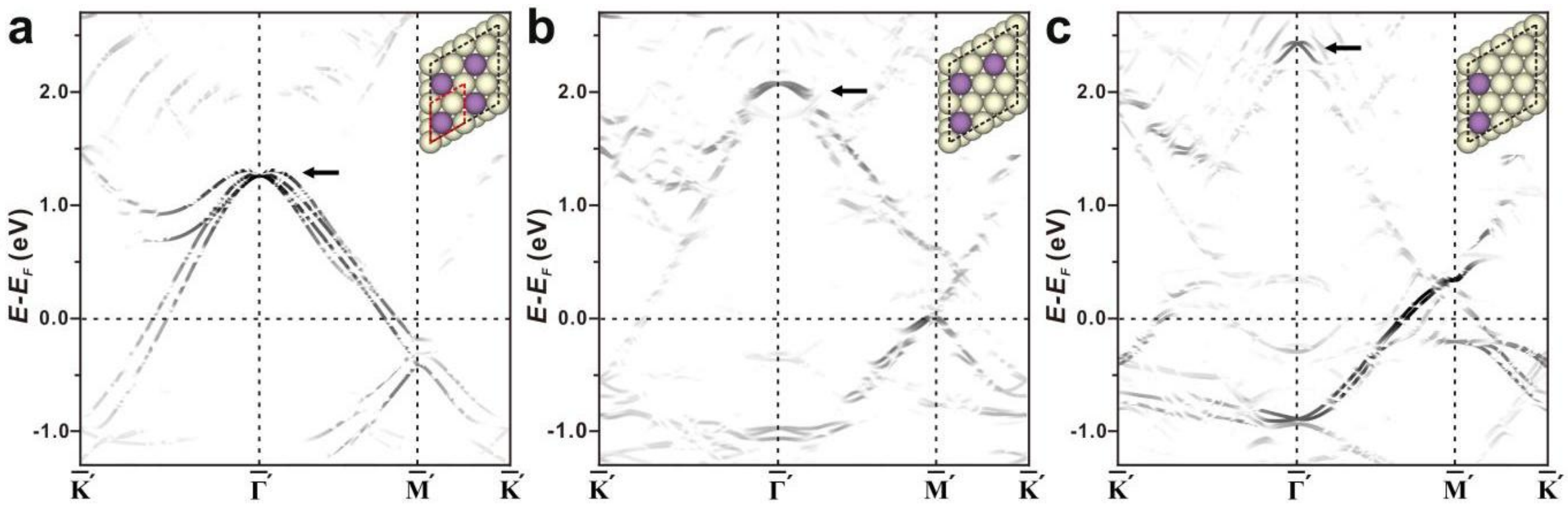

**Figure 4. Orbital decomposed band structure of the $Au_2Sb$ surface alloy with Sb defects.** The structural defects were introduced by subsequently substituting Sb with Au atom. The black symbols indicate the $p_x + p_y$ orbital contributions of Sb in the surface alloy. The $2 \times 2$ slab models are shown at the top right of the corresponding calculation result. (a) perfect $Au_8Sb_4$, (b) $Au_9Sb_3$, (c) $Au_{10}Sb_2$. The crossing point of the Rashba band at $\overline{\Gamma}'$ is marked by a black arrow.

The band structure of the perfect $2 \times 2$ model of the $Au_2Sb$ surface alloy is presented in Figure 4a, showing a clear Rashba-type band splitting as discussed earlier. However, when the Sb atoms in the $2 \times 2$ supercell are gradually substituted with Au atoms one by one to artificially create a “defective” structure, the two well-separated sub-bands gradually undergo an energy-dependent broadening of the linewidth, apart from near the crossing points of the Rashba band. This behavior is typical for non-resonant impurities[28]. Based on our analysis of the superstructure decorated with 3-pointed-star-shaped defects as shown in Figure 1d, the components of Sb and Au within the $8 \times 8$ superstructure can be written as $Au_{135}Sb_{57}$, in which the ratio of Sb is slightly lower than that of the perfect $Au_8Sb_4$ surface alloy model but higher than the $Au_9Sb_3$ defective model considered in Figure 4b. Hence, we deduce that the $S_1$ band of the $8 \times 8$ superstructure is theoretically predicted to exhibit an energy-dependent broadening character but still possesses the Rashba splitting, which is consistent with our experimental observations. Additionally, the calculated band structures in Figures 4a-c demonstrate that the Rashba band, indicated by a black arrow, shifts to a higher energy as the Sb atoms in the $Au_2Sb$ surface alloy are progressively substituted by Au atoms.

## Conclusion

In this work, the Au-Sb alloyed superstructure were prepared under different growth conditions and their atomic and electronic structures were studied utilizing LEED, XPS, (spin-)ARPES, and DFT calculations. From the LEED, STM simulation and the qualitative agreement between the ARPES data and calculated band structure, we find that the Sb atoms alloying on the Au surface can form a moiré superstructure decorated with periodic defects. Apart from the surface alloy deviating from the ideal $\sqrt{3}\times\sqrt{3}R30°$ periodicity, our ARPES measurement reveals a hole-like Rashba-type spin-split band feature $S_1$ with a remarkable broadening character. The good agreement between spin-ARPES results and the theoretically calculated spin texture of the $S_1$ band confirms the helical spin texture of this band. A possible explanation for the broadening character based on the periodic defects playing the role of nonresonant impurities has been proposed and confirmed by our theoretical study of the $Au_2Sb$ surface alloy by introducing defects (impurities) as perturbations. The periodic defects, maintaining the same in-plane inversion symmetry and rotational symmetry as the ideal surface alloy, can properly tune the energy position of the Rashba bands while not ruining the helical spin texture in the Rashba splitting.

## Methods

All sample preparation steps and experiments were performed under ultrahigh vacuum conditions better than $4\times10^{-10}$ mbar. The Au(111) surface was prepared by repeated sputtering and annealing cycles. The quality of the clean surface was confirmed by the existence of sharp diffraction spots observed in low-energy electron diffraction (LEED), corresponding to the well-known Au-herringbone reconstruction, and by the intensity of the surface state, e.g., the Shockley state, confirmed by our ARPES measurements (see Supplementary Figure 1 for details)[24]. The $Au_2Sb$ surface alloy was prepared by depositing a submonolayer of Sb (purity 99.9999%) on clean Au(111) at room temperature (RT). Alternatively, a larger amount of Sb could be deposited and the sample annealed to 400 °C, until a $\sqrt{3}\times\sqrt{3}R30°$ structure could be observed in the LEED. The success of the sample preparation was confirmed by LEED and X-ray photoemission spectroscopy (XPS) spectra of Au 4*f* and Sb 3*d* core levels. Band structure measurements were performed using an aberration-corrected, energy-filtered photoemission electron microscope (EF-PEEM) (NanoESCA $\mathrm{III}$, Scienta Omicron GmbH) equipped with a focused helium discharge lamp predominantly generating He $I$ photons at $h\nu = 21.22$ eV. A pass energy of $E_P = 25$ eV and a 0.5 mm entrance slit to the energy filter were used, yielding nominal energy and momentum resolutions of $\Delta E = 50$ meV and $\Delta k = 0.02$ Å$^{-1}$. The samples were kept at $T \approx 115$ K during the measurements. Spin-resolved ARPES measurements were performed using a monolayer-Au-coated Ir(001) crystal as a spin filter, which exhibited a major spin contribution along $+k_y(-k_y)$ at the scattering energy of 10.95 eV (12.30 eV)[26,29]. The spin filter measured the projection of the spin along $\pm k_y$, which is in-plane and perpendicular to one of the $\overline{\Gamma}-\overline{\mathrm{M}}$ directions ($\pm \mathrm{k_x}$) of the Au(111) surface Brillouin Zone (SBZ). The spin-polarization $P_y$ was calculated from the measurements using[30]:

$$P_y = \frac{I_{y\uparrow} - I_{y\downarrow}}{S\left(I_{y\uparrow} + I_{y\downarrow}\right)}$$

where $I_{y\uparrow}$ and $I_{y\downarrow}$ represent the intensities of the energy surface associated with electron spin along $+k_y$ and $-k_y$, respectively. Preliminary calibration measurements suggested a Sherman function of $S = 0.6$ [30,31].

The density functional theory (DFT) calculations were performed using the Vienna Ab initio Simulation Package (VASP)[32]. The interactions between the valence electrons and ion cores were described using the projector augmented wave method[32,33]. The electron exchange and correlation energy was treated by the generalized gradient approximation with the Perdew-Burke-Ernzerhof functional[34]. The kinetic energy cutoff of the plane-wave basis was set to 400 eV as default. An experimentally confirmed large $8 \times 8$ supercell was not applied for the band structure calculations due to excessive resource consumption. Instead, a simplified $\sqrt{3} \times \sqrt{3}R30°$ model with 22 layers of Au atoms was used, where the bottom 20 layers were fixed as the bulk crystal structure, and the top three layers were relaxed. One monolayer of $Au_2Sb$ atoms was adsorbed on the top side of the Au slab. A vacuum layer thicker than 15 Å was applied to eliminate spurious interactions between adjacent slabs. The first Brillouin zone was sampled with the Γ-centered $15 \times 15 \times 1$ k points. The structures were optimized until the forces on the atoms were less than 20 meV $Å^{-1}$. All the calculated bands are rigidly shifted down by 0.25 eV to match the ARPES data. To study the band structure influenced by periodic defects, $2 \times 2$ supercells were built from the primitive $\sqrt{3} \times \sqrt{3}R30°$ unit cell. In the supercell, Sb atoms were gradually replaced by Au atoms. Spin-orbit coupling was included unless otherwise stated. The $8\sqrt{3} \times 8\sqrt{3}$ superstructure with the bottom 2 layers of Au atoms fixed was applied for STM simulation using the Tersoff-Hamann approach[35].

## Data availability

The data that support the findings of this study are available from the corresponding authors upon request.

## Acknowledgment

This work was partially supported by the Research Council of Norway through its Centres of Excellence funding scheme, Project No. 262633, “QuSpin”. X. S. W. acknowledges support from the Natural Science Foundation of China (NSFC) Grant No. 12174093 and the Fundamental Research Funds for the Central Universities.

## Author contributions

J.B.H. and J.W.W. conceived the project. J.B.H. grew the samples. J.B.H. carried out the ARPES measurements with the assistance from A. C. Å. and J.W.W. J.B.H. and A. C. Å. led the spin-resolved ARPES measurements. J.B.H. and X.S.W. performed the theoretical calculations. J.B.H., X.S.W., A. C. Å. and J.W.W. analyzed the data and wrote the paper. All authors discussed the results and commented on the manuscript.

## Competing interests

The authors declare no competing interests.

## Additional information

**Supplementary information** The online version contains supplementary material available at XXX

**Correspondence** and requests for materials should be addressed to Jinbang Hu and Justin W. Wells.

# Supplemental Information

## Energy-dependent broadening of Rashba-type spin splitting in a $Au_2Sb$ surface alloy with periodic structural defects

Jinbang Hu[1,2 +]; Xiansi Wang[2]; Anna Cecilie Åsland[1]; Justin W. Wells[1,3 +]

[1]*Department of Physics, Norwegian University of Science and Technology (NTNU), NO-7491, Trondheim, Norway.*

[2]*Hunan University, Changsha, 410082, China.*

[3]*Semiconductor Physics, Department of Physics, University of Oslo (UiO), NO-0371 Oslo, Norway.*

**Supplementary Figure 1:** Experimental confirmation of the clean Au(111) surface.

**Supplementary Figure 2:** Characterization of structural phase of the Sb/Au(111) interface.

**Supplementary Figure 3:** ARPES imaging the Rashba-type band feature $S_1$.

**Supplementary Figure 4:** Calculated band structure of the Sb/Au(111) interface with a $\sqrt{3} \times \sqrt{3}$ surface reconstruction.

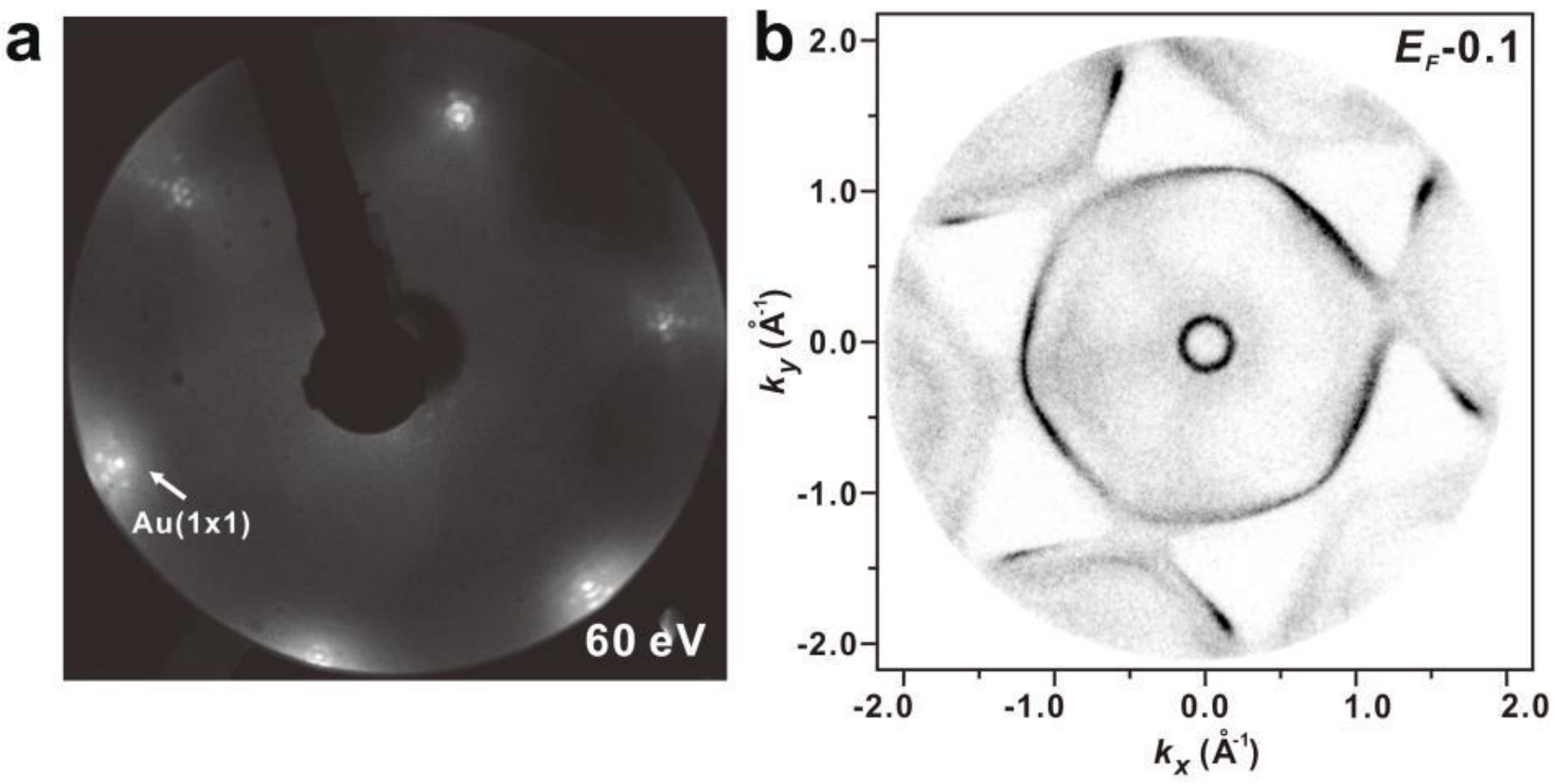

**Figure S1. Experimental confirmation of the clean Au(111) surface.** (a) LEED pattern (60 eV) of the clean Au(111) substrate. (b) ARPES data of the constant energy contour throughout the Au(111) SBZ at $E_b = 0.1$ eV.

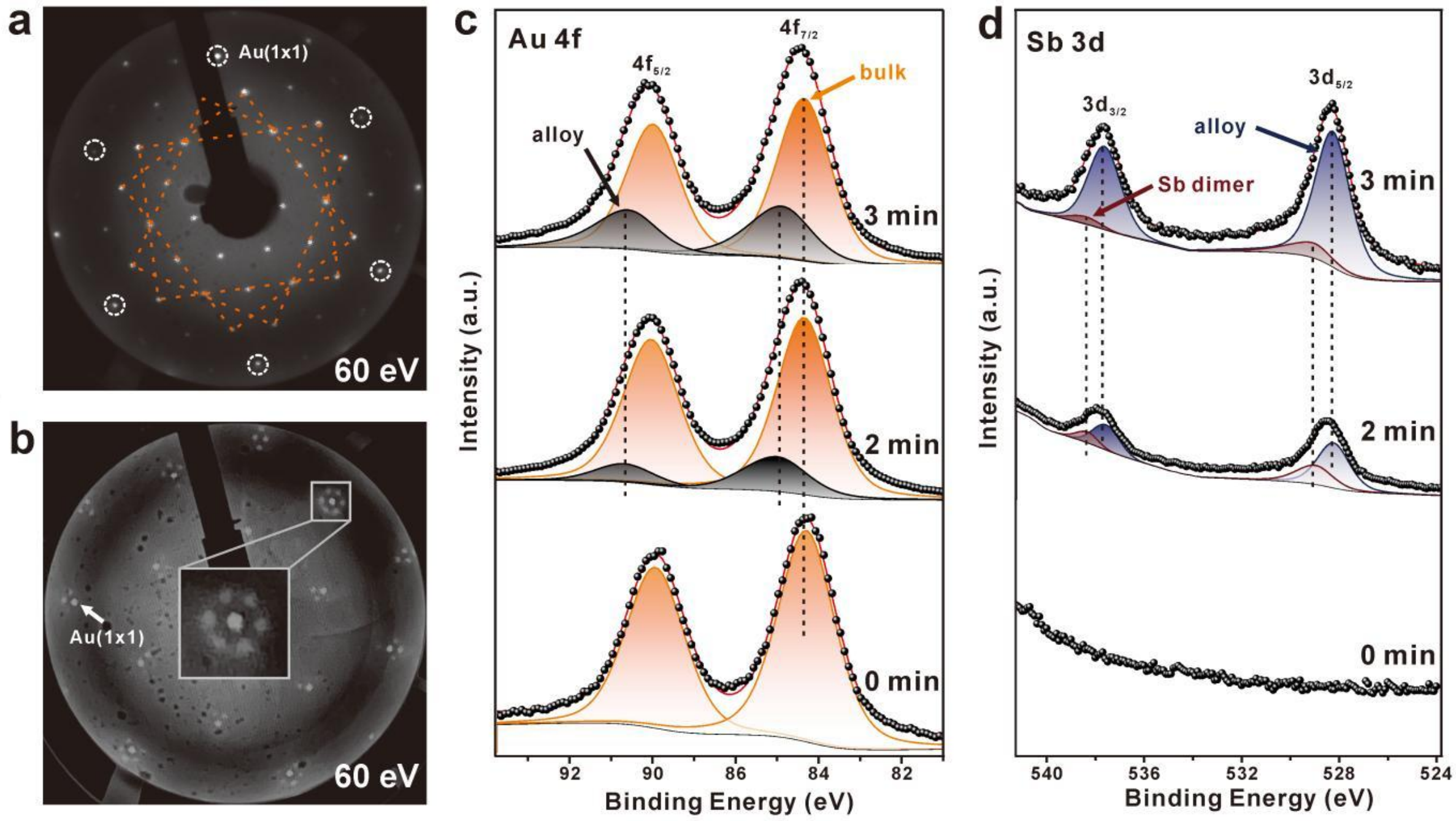

**Figure S2. Characterization of structural phase of the Sb/Au(111) interface.** (a) LEED pattern (60 eV) of a Sb monolayer grown on the Au(111) substrate. The unit cell of 2D Sb is proposed and highlighted by yellow dashed lines. (b) LEED pattern (60 eV) of the $Au_2Sb$ surface alloy formed after 3min of Sb deposition at room temperature. The inset shows the details of the diffraction around the Au(1x1) diffraction spot, presenting an $8 \times 8$ superstructure. (c-d) Evolution of Au 4*f* and Sb 3*d* core level as a function of Sb coverage at room temperature.

In Supplementary Figures 2c-d, the Au 4*f* and Sb 3*d* core levels are presented to monitor the deposition of Sb onto the Au(111) substrate, transitioning from the clean surface to the formation of the $Au_2Sb$ surface alloy at room temperature. The Sb 3*d* signal appears after 1.5 minutes of Sb deposition at room temperature. Concurrently, the Au 4*f* core level shows alterations in both shape and energy. Our peak fitting of the Sb 3*d* signal reveals two components. The high binding energy component might be attributed to Sb dimers formed above the $Au_2Sb$ surface alloy at low coverage, consistent with the STM observation in Ref.[S1] The components at lower binding energies of 537.95 eV and 528.60 eV match well with the Au-Sb alloy signal, as discussed in the main text. Further Sb deposition on the Au surface enhances the Sb 3*d* signal, with a more pronounced component from the surface alloy, while the higher binding energy shoulder diminishes. This change is attributed to the surface phase transition from Sb dimers to the $Au_2Sb$ surface alloy.

The fitting of the Au 4*f* core level indicates that the peak has a higher binding energy surface component, shifted by 0.5 eV from the bulk component. The surface component may result from the interaction between Au and Sb atoms in the surface alloy, and becomes consistently stronger as more Sb is deposited. The LEED pattern in Supplementary Figure 2b shows the $8 \times 8$ superstructure of the $Au_2Sb$ surface alloy, similar to the $8 \times 8$ superstructure formed by the 2D Sb(110) rhombohedral phase on the Au(111) surface upon annealing to 400 °C, as discussed in the main text.

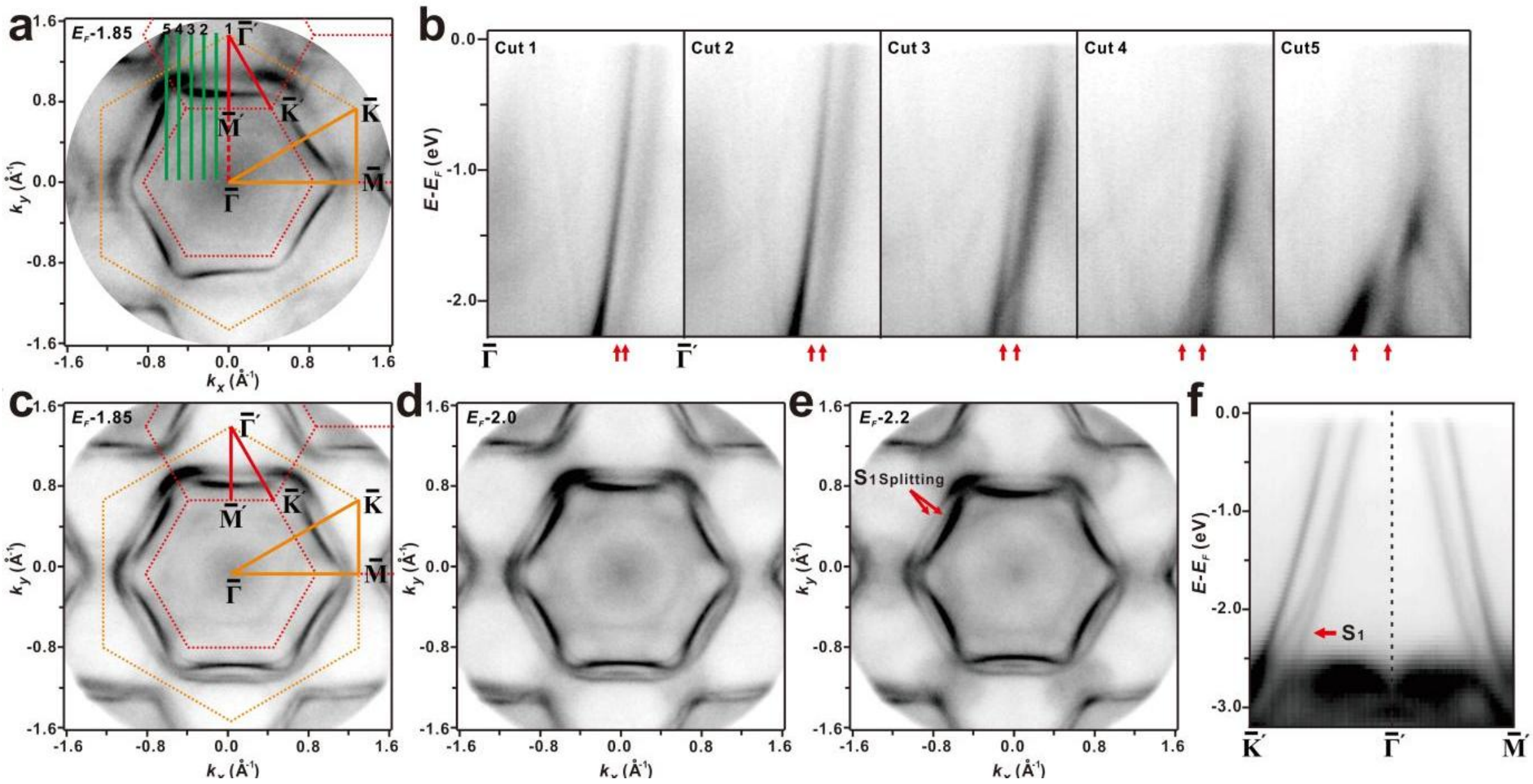

**Figure S3. ARPES imaging the Rashba-type band feature $S_1$.** (a) ARPES data of the $Au_2Sb$ surface alloy showing the constant energy contour throughout the SBZ at 1.85 eV below the Fermi level. (b) Band dispersion along the lines marked as 1-5 in (a) to illustrate the evolution of the splitting of the $S_1$ band feature. (c-e) ARPES data of constant energy contours and (f) band dispersion along the high-symmetry directions of the 2nd SBZ of the $Au_2Sn$ surface alloy, used as a reference to show the clear splitting of the Rashba-type band feature $S_1$.

Supplementary Figures 3a-b show the evolution of the splitting of band feature $S_1$ from the $Au_2Sb$ surface alloy, normal to the $\overline{\Gamma} - \overline{\Gamma}'$ direction. We can clearly see the splitting of the band feature $S_1$ in cuts 2-5. For comparison, Supplementary Figures 3c-f present the electronic structure of the $Au_2Sn$ surface alloy. The sample preparation of $Au_2Sn$ surface alloy followed the procedure described in Ref.[s2]. We found the two splitting branches of the $S_1$ band feature in the $Au_2Sn$ surface alloy to be well separated.

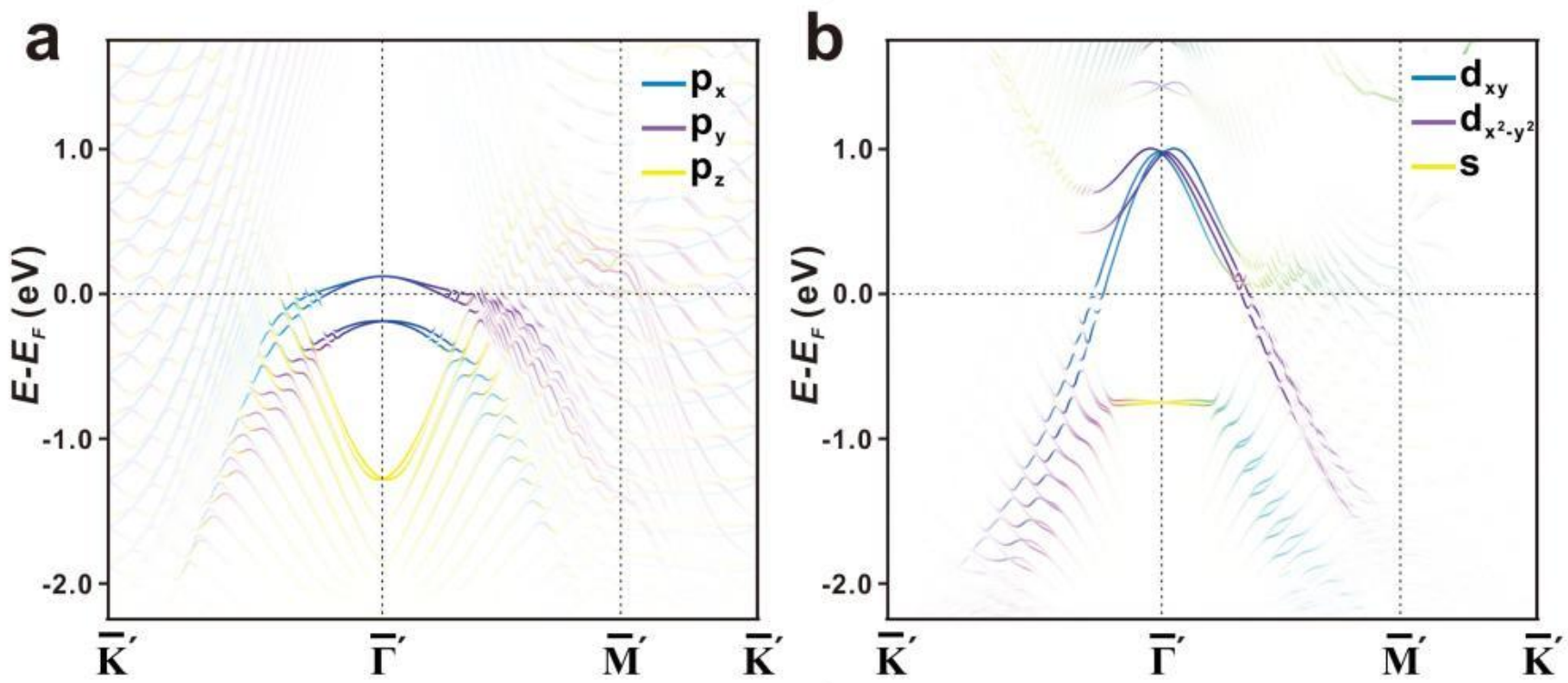

**Figure S4. Calculated band structure of the Sb/Au(111) interface with a $\sqrt{3}\times\sqrt{3}$ surface reconstruction.** (a) Calculated band structure with SOC for the Sb absorbed above the Au(111) with a $\sqrt{3}\times\sqrt{3}$ surface reconstruction. Blue, purple, and yellow symbols indicate contributions from $p_x$, $p_y$, and $p_z$ states projected from the absorbed Sb atom. (b) Calculated band structure with SOC for the model of the $Au_2Sb$ surface alloy shown in Figure 1f. Yellow, blue, and purple symbols indicate contributions from $s$, $d_{xy}$ and $d_{x^2-y^2}$ states of the topmost $Au_2Sb$ layer.

In Supplementary Figure 4a, the surface band projected from the absorbed Sb atom shows a clear band dispersion around the $\overline{\Gamma}'$ point, but its dispersion becomes less pronounced as the binding energy increases. This feature exhibits distinct differences compared to the calculated band structure of the $Au_2Sb$ alloy model and the experimental ARPES data.